\documentclass[sigconf, nonacm]{acmart}
\usepackage{algorithm}
\usepackage{algpseudocode}
\usepackage{graphicx}
\usepackage{tabularx}
\usepackage{textcomp}
\usepackage{xcolor}
\usepackage{graphicx}
\usepackage{epstopdf}
\usepackage{multirow}
\usepackage{hhline}
\usepackage{enumitem}
\usepackage{listings}
\usepackage{anyfontsize}
\usepackage{wrapfig}
\usepackage{footnote}
\usepackage{tablefootnote}
\usepackage{multirow,booktabs}
\usepackage{threeparttable}

\AtBeginDocument{%
  }

\newcommand{\myname}{\textsc{SAAP}}

\begin{document}

\title{\myname{}: An Efficient \underline{S}patial-\underline{A}ware \underline{A}nalytic \underline{P}artitioning Algorithm of VLSI Netlists for Parallel Routing}

\author{Chen Liu$^\dag$, Hongxin Kong$^\ddagger$, Lang Feng$^\dag$$^*$, Wenchao Qian$^\ddagger$, Wuxi Li$^\ddagger$}
\affiliation{%
  \institution{$^\dag$Sun Yat-sen University, China; $^\ddagger$Advanced Micro Devices, Inc., USA}
  \country{\normalsize liuch583@mail2.sysu.edu.cn; konghongxin@outlook.com; fenglang3@mail.sysu.edu.cn; wenchao.qian@amd.com; wuxili@amd.com}
}

\begin{abstract}
As VLSI designs grow in complexity, partitioning is widely adopted to accelerate physical design through parallel computing. However, traditional hypergraph partitioning methods often degrade in performance when applied to 2D layouts due to spatial constraints. For routers with post-placement locations, a spatial-aware partitioning method fully utilizing placement data is preferable. Existing works can only consider soft spatial constraints, leading to a scattered distribution in one partition. We propose \myname, an analytic partitioning algorithm  enforcing hard spatial constraints while efficiently minimizing cut sizes. It includes analytic boundary modeling with regularity-guided simulated annealing and region embedding. Given placed netlists, it generates timing-friendly k-way spatially continuous partitions for parallel routing. Experiments show that it can quickly provide several to dozens of times smaller spatial cut sizes than previous state-of-the-art, with better spatial continuity.
\end{abstract}


\newcommand\blfootnote[1]{%
\begingroup
\renewcommand\thefootnote{}\footnote{#1}%
\addtocounter{footnote}{-1}%
\endgroup
}

\maketitle

\blfootnote{$^*$The corresponding author. $^\S$This work was supported in part by National Natural Science Foundation of China (Grant No. 92473207) and in part by Shenzhen Science and Technology Program (Grant No. JCYJ20241206180301003).}

\vspace{-2ex}

\section{Introduction}
Partitioning has emerged as a cornerstone technique in the realm of VLSI electrical design automation (EDA) flow. By splitting the circuit into partitions and parallelizing computation within each, the runtime of the design process is effectively managed.

Traditionally, the VLSI partitioning challenge centers around hypergraph partitioning, with the goal of achieving a weighted minimum cut. This topic has been the focus of extensive research over many years, exploring a variety of methods such as multilevel algorithms~\cite{Karypis99}~\cite{Springer11}~\cite{Schlag23}, spectral algorithms~\cite{Bustany22}~\cite{Bustany24}~\cite{Sajadinia25}, and neural network-based approaches~\cite{Chen25}, etc. Although partitioning usally performed before placement, incremental hypergraph partitioning with updated weight from placement, target for routing or timing optimization, can not guarantee the spatial continous for updated partition. In addition, many of these techniques overlook the crucial aspect of standard cell placement distribution. Consequently, the cuts generated through topological partitioning may result in larger cut sizes when applied to 2D physical layouts, leading to increased runtime and detrimental degradation of wirelength (timing) for certain nets. 

\begin{figure}[!h]
\centering
\includegraphics[width=1.0\columnwidth]{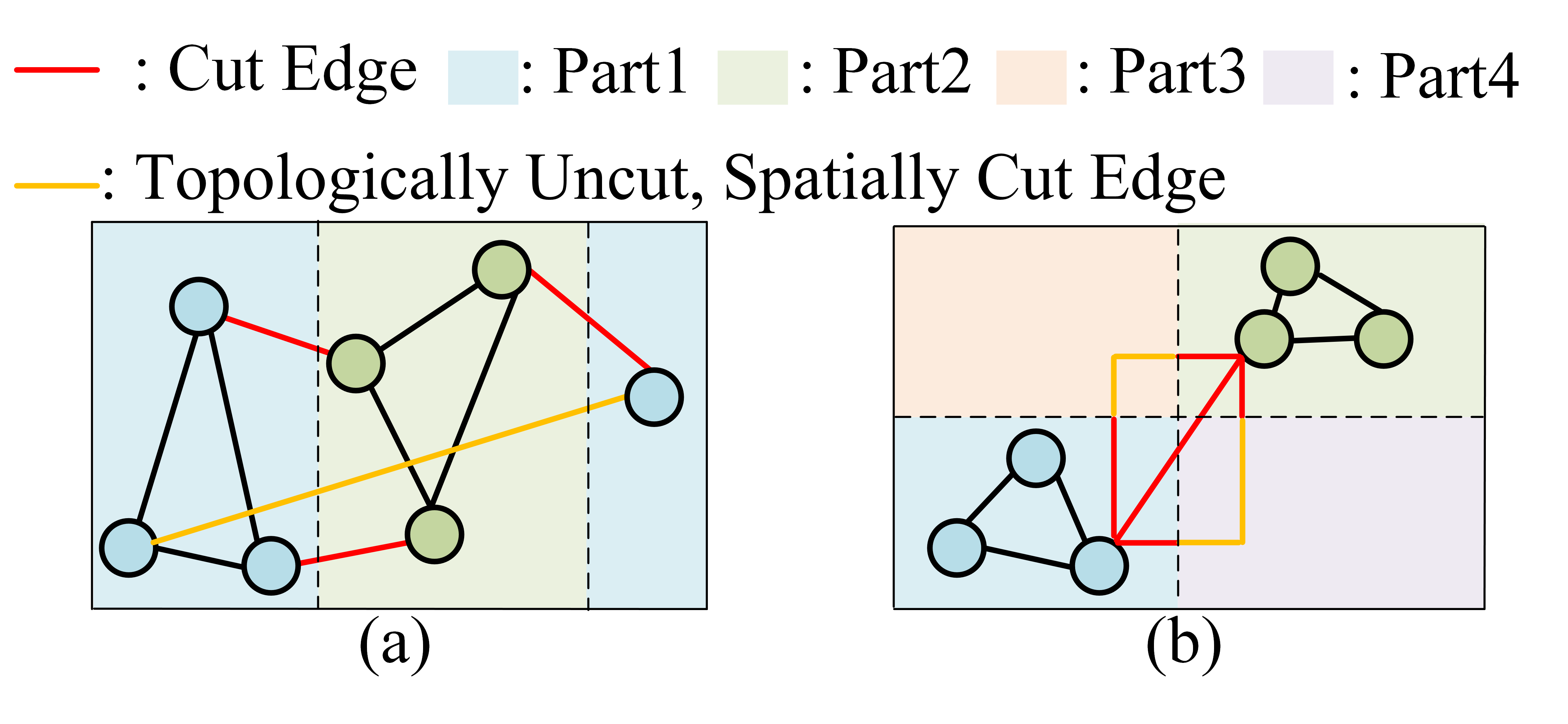}
\vspace{-5ex}
\caption{Mismatch between hypergraph partitioning cut size and spatial-aware cut size.}
\label{fig:problem_explan}
\vspace{-4ex}
\end{figure}

Examples in Figure~\ref{fig:problem_explan} show that unseen cut edges can occur with hypergraph partitioning, and they must be processed to accommodate physical design steps. Nodes in different partitions are marked by different colors. Rectangles in same color belongs to the same partition. In Figure~\ref{fig:problem_explan}(a), the yellow edge, while topologically belonging to the blue partition, crosses the green partition spatially. In Figure~\ref{fig:problem_explan}(b), the red edge indicates a cut edge between partition $1$ and partition $2$. However, to connect the two nodes during the physical design process, routing must pass through either partition $3$ or partition $4$. As a result, the actual spatial-aware cut edges involve three partitions instead of just between partition $1$ and partition $2$. In some modern industrial or academic EDA tools, original nets that cross partition boundaries need to be split into subnets at boundaries to enable fully parallel computing between partitions. However, the behavior also transforms the global routing optimization problem into several local optimization problems in each partition. Minimizing cut size in unweighted graphs or cut weight in weighted graphs is essential for reducing performance degradation caused by partitioning. Unseen spatial cut edges that emerge after hypergraph partitioning can lead to additional subnets and may result in wirelength or timing degradation for certain critical nets. Instead of addressing these spatial cut edges after partitioning, a spatial-aware partitioning algorithm that fully considers the placement of each node is more efficient and conducive to parallel timing optimization and parallel routing.

In this work, we propose \myname{}, an efficient spatial-aware analytic partitioning algorithm that can ensure hard spatial constraints. 
To the best of our knowledge, this is the first hypergraph partitioning work enforcing hard spatial constraints on a 2D plane.
By modeling the partition boundary analytically based on the grid cell-based layout, the accurate weighted cut size on the layout can be efficiently calculated. Besides, a recursive partitioning flow including  regularity-guided simulated annealing and region embedding that leverages the analytic model is proposed. It can separate the layout into multiple spatial continuous partitions. In detail, the contributions of this paper are as follows.

\begin{itemize}
    \item An analytic partition boundary modeling on the 2D layout plane is proposed. This can be leveraged to quickly evaluate the cost after deploying on GPUs.
    \item A recursive partitioning flow leveraging the analytic modeling is proposed. It can efficiently and recursively find a 2-way partition solution for each partition. After region embedding and regularity-guided simulated annealing at each recursion level, a k-way partition solution can be obtained.
    \item Experimental results with multiple state-of-the-art works are conducted. Compared with them under transitional problem formulation as well as spatial-aware partitioning problem, the results indicate that \myname{} can quickly provide several to dozens of times smaller cut sizes than previous state-of-the-art works, with better spatial continuity.
\end{itemize}



\section{Previous Works}
\label{sec:prev}

The classical hypergraph partitioning framework typically follows a multilevel approach~\cite{Karypis99}~\cite{Schlag23}~\cite{Springer11}, which begins by coarsening the hypergraph based on its topological structure. Starting from the coarsest level, the framework employs heuristic optimization methods~\cite{Fiduccia82} to enhance the quality of the partitioning, and it is then mapped back through each level until the partitioning results for the original hypergraph are achieved. 
The quality of the final partitioning is highly sensitive to the quality of the coarsening process. 
Although high-quality coarsening can be achieved by capturing the global topological structure of the hypergraph using techniques based on maximum flow and effective resistance~\cite{Sajadinia25}, these methods introduce additional time overhead and are not directly applicable in scenarios that require the consideration of spatial information.

The introduction of spectral methods has significantly enhanced the analytical modeling of hypergraph partitioning. This approach typically employs the Laplacian matrix of a graph to maintain its global structure~\cite{Chan18}. Commonly, researchers convert the partitioning problem into an eigenvector decomposition~\cite{Bustany22}~\cite{Bustany24}, and then further optimize the results by incorporating input reference solutions. Additionally, some methods utilize existing node embedding techniques~\cite{Perozzi14}~\cite{Grover16}~\cite{Liu24} to generate feature vectors that encapsulate topological information. These feature vectors are subsequently employed in analytical algorithms or heuristics~\cite{Liang24}~\cite{Chen25}. While these spectral frameworks and analytical algorithms have demonstrated effectiveness in achieving the minimum cut, many of the embedding methods focus solely on capturing structural similarity, overlooking node attributes. Furthermore, the computational cost remains prohibitively high for large designs.

Recent studies have explored innovative partitioning tools that account for timing and various physical factors~\cite{Hwang05}~\cite{Liou20}~\cite{Bustany23}. Notably, some research~\cite{Bustany23} has introduced unique constraints based on physical data during the coarsening stage, using the cosine similarity of node eigenvectors as soft guides. The primary goal of this research is to achieve the minimum cut without considering placement information. In addition, the timing analysis in these works does not fully capture the routing that may occur in subsequent stages, leaving room for future improvements.

\section{Overview}
\label{sec:overview}

\subsection{Problem Formulation}
\label{sec:prob}

This work targets a spatial cut boundary that partitions a given placed netlist of a VLSI design into multiple parts. 
The standard cells of each partition need to be spatially continuously located in a part, and the objective is to minimize spatial-aware weight cut size on pre-routed nets. Note that as shown in Figure~\ref{fig:problem_explan}, if a net is crossed by the spatial cut boundary multiple times, the cut size is accumulated.

In detail, the spatial-aware hypergraph partitioning problem of this work is as follows. A hypergraph $H=(V,E)$ includes a set of nodes $V$ and a set of hyperedges $E$. A node $v\in V$ is associated with a weight $w_v$ and a 2D coordinate $p_v=(x_v,y_v)$ as a position. A hyperedge $e\in E$ is a subset of $V$ with weight $w_e$, with a 2D pre-routed tree connection $T_e$ connecting all $v\in e$. Given an integer $k\geq 2$, the hypergraph partitioning problem is to partition $V$ into $k$ disjoint subsets $S=\{V_0, V_1... V_{k-1}\}$ by a spatially continuous boundary $B$ under balance constraint and spatial constraint. For balance constraint, it needs to ensure the node weight of different partitions to be similar. Give a real number $\varepsilon\leq\frac{1}{k}$, and the total weight $W$ of all nodes, the balance constraint is $(\frac{1}{k}-\varepsilon)W\leq\sum_{v\in V_i}w_v\leq(\frac{1}{k}+\varepsilon)W$. For spatial constraint, it requires that the nodes in one partition locate continuously on the 2D layout plane. In detail, for any partition $V_i$, we define the bounding polygon $BP_i$ of $V_i$ as the polygon constructed by some $v\in V_i$ with the smallest area that covers all $v\in V_i$. The spatial constraint requires that there is no overlap between different $BP_i$ on the 2D plane.
Finally, the objective of spatial-aware hypergraph partitioning is the minimal cut size $custsize_H(S)=\sum_e|B\cap T_e|w_e$, where $|B\cap T_e|$ stands for the number of cross points between the spatial boundary $B$ and 2D tree structure $T_e$ of hyperedge $e$.

\subsection{Algorithm Flow}
\label{sec:alg}

\begin{figure}[!h]
\centering
\includegraphics[width=0.9\columnwidth]{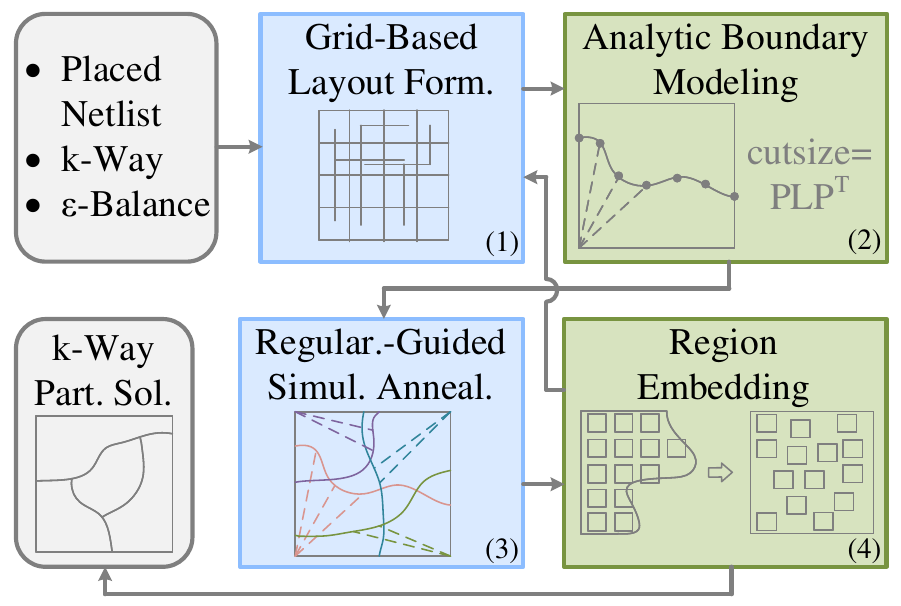}
\caption{The overall flow of \myname.}
\label{fig:flow}
\vspace{-2ex}
\end{figure}

\myname{}'s flow is shown in Figure~\ref{fig:flow}. It uses analytic boundary models based on the grid-based layout. By deploying the cut size calculation on the GPU, the proposed regularity-guided simulated annealing can quickly regress to an optimized solution. The k-way solution is obtained by recursively embedding the 2-way region into new separated rectangle layouts and applying the flow again.
\begin{enumerate}
\item \textbf{Grid-Based Layout Formation:} Given a placed netlist, the number of required partitions $k$, and the balance constraint parameter $\varepsilon$, the first step is to model the layout with grids, such as employing GCells in the router. \myname{} carries out partitioning based on the grid layout. To evaluate the effect of a cut, Steiner tree is generated for all nets as pre-routed trees and then used to assign weights to the edges of the grid.
\item \textbf{Analytic Boundary Modeling:} To speed up the cut size cost calculation, as well as make the boundary spatially continuous, an analytic boundary modeling method is proposed. This approach utilizes polar coordinates for continuity and employs the Laplace matrix for calculating cut size. Additionally, it can be implemented on a GPU for improved speed.
\item \textbf{Regularity-Guided Simulated Annealing:} Using boundary and cut size modeling, a simulated annealing process is performed to optimize the cut size for 2-way partitioning. This process explores various boundaries while ensuring that the boundaries remain regular during updates.
\item \textbf{Region Embedding:} For k-way partitioning, after each 2-way partitioning for each rectangle layout from step (3), each partition is embedded into a new rectangle layout\cite{tutte1963draw,floater2003mean} and repeat steps (1)-(3). 
The process continues recursively through steps (1)–(4) until k-way partitioning is achieved. Once k-way partitioning is completed, the partitions from each embedded rectangle layout are integrated back into the initial layout and presented as the final output.
\end{enumerate}

\vspace{-2ex}
Based on the flow, a k-way partitioning solution with optimized weighted cut size and  spatial continuity is obtained.
\vspace{-2ex}

\section{Algorithm Details}


\subsection{Grid-based Layout Formation}

\begin{figure}[!h]
\centering
\includegraphics[width=0.85\columnwidth]{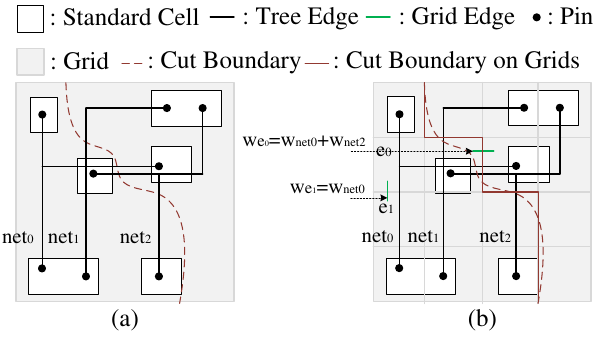}
\caption{Example boundaries on the original and grid-based layouts. (For clarity, only two grid edges are shown.)}
\label{fig:grid}
\vspace{-2ex}
\end{figure}

Traditionally, the nodes of the hypergraph represent standard cells, while the hyperedges correspond to the nets. An example of partitioning is illustrated in Figure~\ref{fig:grid}(a). 
However, when we have the placement information and to better accommodate the routing problem, a grid-based hypergraph is adopted to manage the minimum resolution for the partitioning. For instance, when cells are placed in a rectangular layout like in Figure~\ref{fig:grid}(a), the area is divided into multiple grids, as shown in Figure~\ref{fig:grid}(b). Each pair of adjacent grids, whether positioned horizontally or vertically, has a weighted edge, exemplified by edges \(e_0\) and \(e_1\) in Figure~\ref{fig:grid}(b). In our experiments, GCells from the router are used to construct the grid graph.

Based on the grid-based layout, the partitioning task on the initial hypergraph can be mapped to partitioning on the $n\times n$ grid graph $G=(V,E,\{w_v\},\{w_e\})$ , with grids as the nodes $V$ and grid edges $E$ as the hyperedges.  The weight $w_v$ of each grid node $v\in V$ represents the number of pins in the grid, while the weight $w_e$ of each grid edge $e\in E$ is the cumulative weight of all the nets that pass through the boundary of the grid, based on the routing estimated by Steiner trees.  FLUTE~\cite{Chu08} algorithm is applied in the experiments to generate tree solutions for nets. For example, the tree generation solutions of nets $net_0$ to $net_2$ in Figure~\ref{fig:grid}(b) are shown as black lines. Grid edge $e_0$ has weight $w_{e_0}=w_{net_0}+w_{net_2}$, while grid edge $e_1$ has weight $w_{e_1}=w_{net_0}$. 
The cut for the grid graph is rectilinear and passes through the grid boundaries. For example, a curved cut shown in Figure~\ref{fig:grid}(a) corresponds to the rectilinear cut boundary that traverses the grid borders in Figure~\ref{fig:grid}(b). If each net's weight is 1, the cut size is 4 in the example.

\subsection{Analytic Boundary Modeling}
\label{sec:bound}
\begin{figure}[!h]
\centering
\includegraphics[width=1.0\columnwidth]{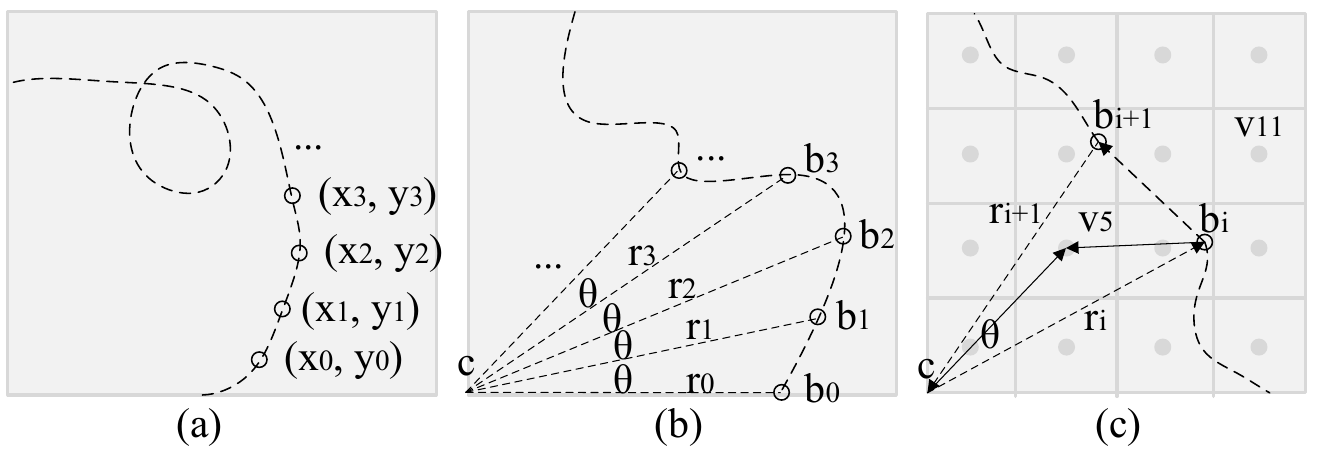}
\caption{The illustration of the boundary modeling.}
\label{fig:bound}
\vspace{-2ex}
\end{figure}

The next step involves modeling the cut boundary and determining the cut cost associated with this boundary. To enhance efficiency, this process is modeled analytically.
In this work, k-way partitioning is executed recursively through 2-way partitioning, meaning that the boundaries for two partitions are considered for each recursion level. For spatial-aware partitioning, the boundary must also be represented as a continuous curve. One straightforward approach is to model the curve using multiple discrete points, with each nearest pair of points connected by a straight line. Each point's coordinates, represented as \((x_i, y_i)\), become decision variables. When a large number of points are used, the connections between them can approximate a curve, as illustrated in Figure~\ref{fig:bound}(a).
However, this method can lead to the illegal boundary, as shown in the example, which violates spatial constraints. Although additional constraints can be implemented to regulate the related positions of the points, doing so can significantly increase the complexity of the model.

In this work, we propose a boundary modeling that intrinsically avoids the illegal boundaries. Shown as the example in Figure~\ref{fig:bound}(b), the boundary is also modeled as multiple connected points $b_i$, where the position of each point is represented by the polar coordinate. 
First, one corner of the rectangle layout is selected as the origin $o$, such as the bottom-left corner in the example. 
Then, the $\frac{\pi}{2}$ degree corner is separated into $\frac{\pi}{2\theta}$ angles, with each angle $\theta$ degree. There is a radius $r_i$ associated with each angle, and the boundary point $b_i$ is at coordinate $(r_icos(i\theta), r_isin(i\theta))$. The decision variables are all $r_i$. 
The boundary is constructed by connecting pairs of nearest boundary points $b_i$ and $b_{i+1}$ with straight lines.
By modeling under the polar coordinate, the related position of each $b_i$ is fixed, and no circle can be formed given $r_i\geq0$.

Next, given a cut boundary, the partitioning solution of each grid needs to be modeled. This is approximately modeled by finding out if a center node of a grid is covered by any triangle $\triangle ob_ib_{i+1}$ or not. 
For example, in Figure~\ref{fig:grid}(c), a grid center node $v_5$ is covered by $\triangle ob_ib_{i+1}$, while $v_{11}$ is not, so $v_5$ and $v_{11}$ are at different partitions.
This can be analytically modeled by the cross product. An indicator function $\mathbf{1}[\cdot]$ ($1$ if the condition holds, and $0$ otherwise) is applied to map the cross product results into binary. This is supported by many machine learning toolkit for GPU parallel computing. The covering status of a grid center $v$ under $\triangle ob_i b_{i+1}$ can be modeled as follows, with 1 indicating $v$ is covered by this triangle.
\begin{equation}
Cover_{v, \triangle ob_i b_{i+1}} =
\mathbf{1}\Big[
(\overrightarrow{ob_i} \times \overrightarrow{ov}) 
(\overrightarrow{b_i b_{i+1}} \times \overrightarrow{b_i v}) 
(\overrightarrow{b_{i+1}o } \times \overrightarrow{b_{i+1} v}) \ge 0
\Big]
\end{equation}

 For each grid center $v$, the partition $p_v$ (0 or 1) it belongs is:
 \begin{equation}
 p_v=
\mathbf{1}\Big[
\sum_{i}Cover_{v,\Delta ob_ib_{i+1}} > 0\Big]
 \end{equation}

Then, a grid partition vector 
$P = [p_{v_0},\, p_{v_1},\dots,\, p_{v_{n^2-1}}]$ can be obtained, 
where $p_{v_i}$ corresponds to the partition of the $i$-th grid. 

Next, the weighted cut size, which is the cost, needs to be calculated. 
Let $L$ be the Laplacian matrix~\cite{Bustany22} of the grid graph, which contains the topology information of the graph, then the cutsize corresponding to $P$ is given as follows.
 \begin{equation}
 cutsize= 
 \mathbf
\, P LP ^\top
 \end{equation}
 
Let $D_w$ denote the node-weight matrix of the grid graph, which is a diagonal matrix whose diagonal entries encode the weight of each grid (i.e., the number of standard cells contained in it). $W_v$ is the total weight of the grids. The balance constraint is as follows.
 \begin{equation}
 UB=
  \mathbf
  |2(P D_w P^\top) - W_v|
  \end{equation}

Finally, the cost corresponding to $P$  can be expressed as the weighted sum of the normalized values of the(3)(4) :
 \begin{equation}
 cost=
 \alpha_c \times \frac{cutsize}{W_e} + \alpha_b \times  \frac{UB}{W_v}
\end{equation}

$W_e$ is the total weight of grid edges, while $\alpha_c$ and $\alpha_b$ are custom parameters.
Note that the proposed cost calculation can be deployed on GPUs by adopting machine learning toolkits.
In this case, given a grid graph $G$, an origin $o$ and a radius set $R=\{r_0,r_1...\}$, the cost can be obtained, and this process is named $cost=Cost(R,G,o)$.

\subsection{Regularity-Guided Simulated Annealing}

In this subsection, the 2-way analytic spatial-aware partitioning algorithm based on grid graph is proposed as Algorithm~\ref{alg:2way}. The proposed algorithm is based on simulated annealing, with the analytic cut boundary modeling, cost calculation method, and the boundary regularity enforcement method.

For Algorithm~\ref{alg:2way}, the inputs are the grid graph $G$, the number of angles $m$, and the boundary origin $o$. 
The $\theta$ is firstly calculated on Line 1. Next, the initial variables to form the radius vector $R^0$ of the cut boundary are initialized on Line 2-3. 
Note that, instead of directly initializing $\{r^0_0,r^0_1,...,r^0_m\}$ for $R^0_0$, a base number $r^0_0$ and the differences $\{\Delta r^0_1, \Delta r^0_2, ...,\Delta r^0_m\}$ are initialized (Line 2), and $R^0_0$ is obtained by cumulatively adding the difference $\Delta r^0_i$ with the former radius $r^0_{i-1}$ for $r^0_i$ (Line 3). The reason will be elaborated later in this subsection.
The temperature $T$ and the iteration $i$ of simulated annealing are initialized on Line 4. 
The cost is calculated in Line 5, according to Equation (5) in Section~\ref{sec:bound}.
The best cost and boundary are recorded as the initial solution for simulated annealing (Line 6).

\begin{algorithm}[!h]
\caption{2-Way Grid-Based Spatial-Aware Partitioning} 
\raggedright
{\bf Function:}  
2Way\_Spatial\_Part\\
{\bf Input:}  
Grid Graph $G=(V,E,\{w_v\},\{w_e\})$; 
\# of Angles $m$; Boundary Origin $o$;

{\bf Output:}  
2-Way Partitioning Solution $\{V_0,V_1\}$.
\begin{algorithmic}[1]
\State $\theta \gets \frac{\pi}{2m}$
\State Initialize $r^0_0\gets 0$ and $\Delta R^0\gets \{\Delta r^0_1, \Delta r^0_2, ...,\Delta r^0_m\}$
\State$R^0\gets \{r^0_0,r^0_1,...,r^0_m\}=\{r^0_0,r^0_0 + \Delta r^0_1,r^0_1+\Delta r^0_2,...,r^0_{m-1}+\Delta r^0_m\}$
\State $T\gets T_{init}; k\gets 0$
\State $cost_{prev}\gets Cost(R^0,G,o)$
\State $cost^{best}\gets cost^{prev}$; ${r^{best}_{0}\gets r_0; \Delta R^{best}}\gets  \Delta R^0$
\While{$T\geq T_{limit}$}
    \State $r^{new}_0,\Delta R^{new} \gets Perturb(r^{k}_0, \Delta R^k, T)$
    \State $R^{new}\gets \{r^{new}_0,...,r^{new}_i,...\}=\{r^{new}_0,...,r^{new}_{i-1} + \Delta r^{new}_i,...\}$
    \State $cost\gets Cost(R_{new},G,o)$
    \If{$cost<cost^{prev}$}
        \State $r^{k+1}_0,\Delta R^{k+1}\gets r^{new}_0,\Delta R^{new}$
        \If{$cost<cost^{best}$}
        \State  $r^{best}_0,\Delta R^{best},cost^{best}\gets r^{new}_0,\Delta R^{new},cost$
        \EndIf
    \ElsIf{$e^{\frac{(cost^{prev}-cost)}{T}}>Rand(0,1)$}
        \State  $r^{k+1}_0,\Delta R^{k+1}\gets r^{new}_0,\Delta R^{new}$
    \EndIf
    \State $cost^{prev} \gets cost$ 
    \State $T\gets \gamma T$
\EndWhile
\State Calculate the partition vector $P$ under $R^{best}$ by $r^{best}_0, \Delta R^{best}$
\State $V_0\gets {v_i}$ for $p_{v_i}==0$, $V_1\gets {v_j}$ for $p_{v_j}==1$
\State Return $\{V_0, V_1\}$
\end{algorithmic}
\label{alg:2way}
\end{algorithm}

The iterations of simulated annealing are performed within Lines 7-21.
The first step is to perturb the previous k-th iteration's base $r^{k}_0$ and differences $\Delta r^k_i$, according to the current temperature $T$ (Line 8).
This process aims to identify the neighbors of the decision variables from the previous iteration. The perturbation must adhere to certain properties:
1. For a standard cut boundary in the layout for parallel routing, the boundary should be regular and smooth, ensuring that the difference between nearby radius is limited.
2. The perturbation must be random.
3. To generate a neighbor, the range of the perturbation should not be too large.
4. Each radius must differ from the others to ensure diversity.
To achieve these goals, the proposed perturbation method is as follows.
\vspace{-1ex}
\begin{align}
\Delta r^{new}_{i}=\beta\frac{T}{T_{init}}sin(\Delta r^k_i+\mathcal{N}(0,\,\sigma^{2}))
\end{align}
Properties 2 and 3 are satisfied by $\mathcal{N}(0,\,\sigma^{2})$ and $sin()$, respectively. Meanwhile, since $sin()$ is a continuous function with a smooth curve, by properly setting $\sigma$, $\Delta r^{new}_i$ can be smooth (property 1).
Note that if $\Delta r^{new}_i$ is added to previous $r^{k}_i$ directly, the perturbation might be periodic, which is not the case of a normal cut boundary, and this violates property 4. Therefore, the $r^{new}_i$ is added cumulatively as $\{r^{new}_0,...,r^{new}_i,...\}=\{r^{new}_0,...,r^{new}_{i-1} + \Delta r^{new}_i,...\}$ for addressing the periodic problem.
For the previous base radius $r^{k}_0$, it is perturbed randomly by generating a ranged difference to add to it. Then, the updated radius vector $R^{new}$ is obtained (Lines 8-9), and the current cost is calculated (Line 10).


After this, the ordinary simulate annealing process is performed (Lines 11-20).
Finally, the best partition solution is returned (Lines 22-24).
Note that the input boundary origin $o$ is tried for all 4 corners of the layout, and the best boundary with the smallest cost is kept.

\subsection{Region Embedding for k-Way Partitioning}

\begin{figure}[!h]
\vspace{-3ex}
\centering
\includegraphics[width=0.9\columnwidth]{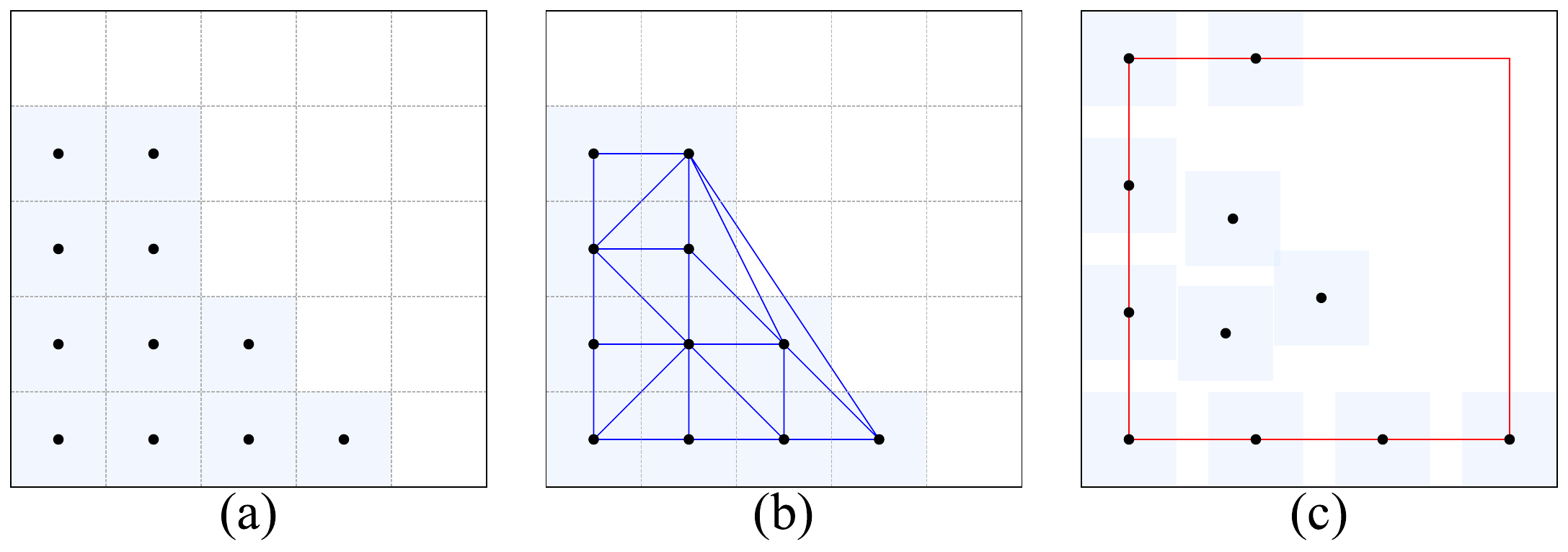}
\vspace{-3ex}
\caption{An example of embedding a partition region to a rectangle layout.}
\label{fig:embed}
\vspace{-2ex}
\end{figure}

Algorithm~\ref{alg:2way} can provide 2-way partitioning solutions. For k-way partitioning, Algorithm~\ref{alg:2way} can be performed recursively for each partition from the previous solution. To match the proposed boundary model, the grids of a partition with an irregular shape on the 2D plane (Figure~\ref{fig:embed}(a)) need to be embedded in a rectangle layout while maintaining the relationship of grids.
Here we use a discrete Harmonic interpolation approach~\cite{floater2003mean}. Due to space limitation, the basic idea of the approach is briefly described as follows.

First, the centers of the grids are connected into a triangular mesh by Delaunay triangulation~\cite{tutte1963draw}, such as Figures~\ref{fig:embed}(a) to (b). Edges that belong to exactly one triangle are marked as boundary edges, and their endpoints are identified as boundary vertices. These boundary vertices are normalized to the unit square and then stretched to the desired rectangle width and height, as shown in Figure~\ref{fig:embed}(c), yielding fixed Dirichlet positions. Treating these positions as boundary conditions, the embedded coordinates of all interior vertices are obtained by solving a discrete Laplace equation whose system matrix is the graph Laplacian of the triangular mesh. This embeds grids with irregular boundary partitioning into a rectangle.

By embedding the partitioned region in a rectangle layout again, it allows the use of a 2-way partitioning algorithm iteratively to implement k-way partitioning.

\section{Experiments}
\label{sec:exp}

\begin{table*}[t]
\caption{Comparisons unweighted and weighted partitioning of different works.}
\label{tab:result}
\centering
\small
\setlength{\tabcolsep}{3pt}

\resizebox{\textwidth}{!}{
\renewcommand{\arraystretch}{0.55}
\begin{tabular}{c | c c c| c c c c| c c c c |c c c c}
\hline
\multirow{25}{*}{\rotatebox[origin=c]{90}{Unweighted}} & \multirow{2}{*}{Testcase} & \multirow{2}{*}{\# Net} & \multirow{2}{*}{\# Cell} 
& \multicolumn{4}{c|}{Cut Size ($\varepsilon = 0.1$)} 
& \multicolumn{4}{c|}{\# Fragments} 
& \multicolumn{4}{c}{Runtime (s)} \\ 
\cline{5-16}

& & & 
 & hMETIS & SHyPar & TritonPart & \myname{}
& hMETIS & SHyPar & TritonPart & \myname{}
 & hMETIS & SHyPar & TritonPart & \myname{}\\
 \cline{2-16}

 & ispd18\_test1 & 3K & 9K & 4K & 4K & 2K & 484 & 166 & 152 & 107 & 2 & 1 & 23 & 27 & 6 \\
 & ispd18\_test2 & 36K & 36K & 13K & 13K & 13K & 2K & 363 & 359 & 359 & 2 & 3 & 54 & 22 & 8 \\
 & ispd18\_test3 & 36K & 36K & 17K & 17K & 18K & 2K & 501 & 501 & 513 & 2 & 3 & 54 & 21 & 10 \\
 & ispd18\_test4 & 72K & 72K & 374K & 96K & 398K & 3K & 8K & 2K & 9K & 2 & 7 & 122 & 40 & 18 \\
 & ispd18\_test5 & 72K & 72K & 99K & 360K & 232K & 4K & 2K & 7K & 5K & 2 & 7 & 136 & 34 & 16 \\
 & ispd18\_test6 & 108K & 108K & 379K & 931K & 909K & 4K & 8K & 21K & 21K & 2 & 12 & 203 & 49 & 22 \\
 & ispd18\_test7 & 180K & 180K & 103K & 942K & 825K & 7K & 3K & 21K & 19K & 2 & 25 & 661 & 83 & 38 \\
 & ispd18\_test8 & 180K & 192K & 2M & 644K & 899K & 7K & 60K & 18K & 28K & 2 & 24 & 749 & 75 & 48 \\
 & ispd18\_test9 & 179K & 193K & 902K & 262K & 613K & 5K & 23K & 7K & 16K & 2 & 25 & 810 & 73 & 26 \\
 & ispd19\_test1 & 3K & 9K & 9K & 7K & 4K & 520 & 366 & 289 & 187 & 2 & 1 & 23 & 17 & 6 \\
 & ispd19\_test2 & 72K & 72K & 440K & 594K & 303K & 3K & 10K & 8K & 7K & 2 & 7 & 140 & 39 & 16 \\
 & ispd19\_test3 & 9K & 8K & 2K & 1K & 2K & 563 & 458 & 283 & 455 & 2 & 1 & 20 & 10 & 6 \\
 & ispd19\_test4 & 151K & 146K & 10K & 9K & 5K & 4K & 39 & 49 & 15 & 2 & 8 & 310 & 31 & 34 \\
 & ispd19\_test5 & 29K & 29K & 7K & 7K & 8K & 894 & 75 & 75 & 86 & 2 & 18 & 46 & 12 & 7 \\
 & ispd19\_test6 & 180K & 180K & 940K & 373K & 647K & 7K & 29K & 12K & 19K & 2 & 23 & 780 & 80 & 47 \\
 & ispd19\_test7 & 359K & 360K & 275K & 275K & 276K & 7K & 8K & 8K & 8K & 2 & 40 & 681 & 160 & 82 \\
 & ispd19\_test8 & 537K & 540K & 1M & 104K & 104K & 11K & 35K & 3K & 3K & 2 & 60 & 1350 & 267 & 105 \\
 & ispd19\_test9 & 895K & 899K & 2M & 2M & 142K & 14K & 55K & 49K & 36K & 2 & 103 & 2113 & 540 & 153 \\
\cline{2-16}
&  \multicolumn{3}{c|}{Norm. Sum. Ratio}  & 112.0  & 84.8  & 66.3  & 1.0  & 6751.8  & 4402.3  & 4793.5  & 1.0  & 0.6  & 12.8  & 2.4  & 1.0 \\
\cline{2-16}

\hline
\end{tabular}
} 

\vspace{1ex}

\resizebox{\textwidth}{!}{
\renewcommand{\arraystretch}{0.6}
\begin{tabular}{c | c c c c| c c c | c c c |c c c|c c c}
\hline
\multirow{25}{*}{\rotatebox[origin=c]{90}{Weighted}} &\multirow{2}{*}{Testcase} & \multirow{2}{*}{\# Net} & \multirow{2}{*}{\# Cell} & \multirow{2}{*}{\# Way}
& \multicolumn{3}{c|}{Cut Size ($\varepsilon = 0.1$)} 
& \multicolumn{3}{c|}{\# Critical Nets Crosses} 
& \multicolumn{3}{c|}{\# Fragments} 
& \multicolumn{3}{c}{Runtime (s)} \\ 
\cline{6-17}

& & & & 
 & hMETIS  & TritonPart & \myname{}
& hMETIS  & TritonPart & \myname{}
& hMETIS  & TritonPart & \myname{}
 & hMETIS  & TritonPart & \myname{}\\ \cline{2-17}
\cline{2-17}
 & ariane & 124K & 120K & 2 & 16M & 2M & 684K & 583K & 557K & 25K & 2K & 1K & 2 & 19 & 51 & 27 \\
 & mempool\_tile & 135K & 133K & 2 & 23M & 2M & 687K & 708K & 713K & 22K & 4K & 2K & 2 & 16 & 63 & 15 \\
 & NV\_NVDLA & 199K & 195K & 2 & 12M & 1M & 973K & 417K & 425K & 40K & 3K & 4K & 2 & 23 & 60 & 61 \\
 & bsg\_chip & 736K & 706K & 2 & 436M & 475M & 3M & 11M & 11M & 77K & 13K & 27K & 2 & 110 & 375 & 101 \\
 & mempool\_group & 3M & 3M & 2 & 538M & 481M & 18M & 42M & 42M & 619K & 27K & 30K & 2 & 766 & 1028 & 1090 \\\cline{2-17}
 &  \multicolumn{4}{c|}{Norm. Sum. Ratio}  & 43.9  & 41.1  & 1.0  & 68.8  & 70.9  & 1.0  & 4871.1  & 6417.6  & 1.0  & 0.7  & 1.2  & 1.0  \\\cline{2-17}
 & ariane & 124K & 120K & 4 & 28M & 3M & 2M & 1M & 809K & 58K & 5K & 4K & 4 & 33 & 278 & 66 \\
 & mempool\_tile & 135K & 133K & 4 & 27M & 23M & 2M & 919K & 892K & 58K & 5K & 4K & 4 & 33 & 456 & 46 \\
 & NV\_NVDLA & 199K & 195K & 4 & 21M & 22M & 2M & 758K & 788K & 68K & 5K & 6K & 4 & 42 & 510 & 195 \\
 & bsg\_chip & 736K & 706K & 4 & 556M & 608M & 10M & 15M & 16M & 248K & 36K & 46K & 4 & 205 & 1849 & 293 \\
 & mempool\_group & 3M & 3M & 4 & 673M & 768M & 68M & 68M & 86M & 1M & 66K & 65K & 4 & 1333 & 3817 & 1590 \\\cline{2-17}
 &  \multicolumn{4}{c|}{Norm. Sum. Ratio}  & 15.7  & 17.1  & 1.0  & 46.0  & 56.1  & 1.0  & 5897.6  & 6252.5  & 1.0  & 0.8  & 3.2  & 1.0  \\\cline{2-17}
 & ariane & 124K & 120K & 8 & 45M & 43M & 4M & 2M & 2M & 130K & 8K & 5K & 8 & 42 & 1005 & 126 \\
 & mempool\_tile & 135K & 133K & 8 & 33M & 32M & 23M & 1M & 1M & 95K & 4K & 5K & 8 & 51 & 1679 & 82 \\
 & NV\_NVDLA & 199K & 195K & 8 & 56M & 50M & 6M & 2M & 2M & 190K & 7K & 6K & 8 & 61 & 4940 & 380 \\
 & bsg\_chip & 736K & 706K & 8 & 686M & 657M & 17M & 17M & 16M & 394K & 42K & 48K & 8 & 275 & 6969 & 522 \\
 & mempool\_group & 3M & 3M & 8 & 1009M & - & 99M & 100M & - & 3M & 88K & - & 8 & 1906 & (>5h) & 2122 \\\cline{2-17}
 &  \multicolumn{4}{c|}{Norm. Sum. Ratio (Cases with Full Results)} & 16.6  & 15.8  & 1.0  & 27.1  & 25.4  & 1.0  & 1920.4  & 2008.7  & 1.0  & 0.4  & 13.1  & 1.0  \\

\hline
\end{tabular}
} 

\end{table*}

\subsection{Experimental Setup}

The experiments are performed on a platform with Intel Xeon CPU, NVIDIA GeForce RTX 3090 GPU and 32 GB of DDR4 memory. The analytic modeling is deploy on GPU by PyTorch toolkit~\cite{pytorch}.
The benchmarks are selected from the contests of ICCAD'19~\cite{iccad19} (including testcases from \textit{ISPD'18} and \textit{ISPD'19}), and ISPD'25~\cite{ispd25}. 
The compared algorithms include an advanced partitioning tool hMETIS~\cite{Karypis99}, a recent work SHyPar~\cite{Sajadinia25}, and a partitioning tool TritonPart~\cite{Bustany23} with soft spatial constraints enabled.
The experiments are conducted under netlists with unweighted edges and weighted edges separately. For unweighted edges, ISPD18 and ISPD19 testcases are used. It conducts 2-way partitioning for testing the ability of the partitioning algorithm to provide a small cut size while maintaining the spatial continuity. 
For weighted edges, the weights are set based on the slacks, so only ISPD'25 contest benchmark is used since it contains the timing library. The weights are set according to different ranges of slacks, with a larger weight for a smaller (or negative) slack. Besides, SHyPar is excluded as it does not support weighted edges. It conducts 2, 4, and 8-way partitioning for testing if the proposed partitioning algorithm can provide timing-friendly cut boundaries while maintaining the spatial continuity.

\subsection{Comparisons with Previous Works}

The unweighted partitioning is first conducted, and the results are in Table~\ref{tab:result}, where ``Norm. Sum. Ratio'' is the normalized ratio to \myname{} of the sum of each column.
It indicates that \myname{} has dozens of times smaller cut size compared with all previous works, due to the fact that the cut size in the problem formulation (Section~\ref{sec:prob}) is spatially considered. Topological small cut size does not indicate a spatially small one, as shown in Figure~\ref{fig:problem_explan}. This result shows the effectiveness of \myname{} on the core metric.
Meanwhile, the number of fragments are reported, which are scattered regions of partitions. It indicates that for 2-way partitioning, there are exactly 2 regions by \myname{}, proving that the spatial constraint is satisfied. In contrast, there are many scattered small regions for other algorithms. A detailed example from hMETIS can be illustrated in Figure~\ref{fig:frag}(a), where the two colors represent the Gcells of two partitions. It shows that although most of the Gcells are spatially continuous, there are some Gcells scattering.

Finally, the runtime is evaluated. The proposed analytic algorithm has several times higher speed compared with previous works ShyPar and TritonPart, showing the efficiency. Although hMETIS has quicker speed, its partitioning quality on spatial cut size or continuity are much worse. In this case, the time overhead of the proposed approach compared with hMETIS is reasonable.

Next, the partitioning on the weighted graph is also tested in Table~\ref{tab:result}. The weights of nets (and grid edges) are set according to the slacks from timing analysis. Note that the normalized ratio is calculated without the largest testcase where TritonPart fails to finish within the set time limit (5 hours). The results also indicate similar conclusions on the spatial (weighted) cut size, spatial continuity, and runtime as the unweighted experiments, which prove the effectiveness and efficiency of \myname{}. 
More importantly, the number of crosses on the critical net with negative slacks is calculated. It shows that \myname{} crosses critical nets much fewer times compared with previous works, when considering the slacks by setting the weights. This further proves the timing-friendly property of \myname{} considering spatial connections of the netlists.

\subsection{Insight Analysis}

\begin{figure}[!h]
\centering
\includegraphics[width=0.9\columnwidth]{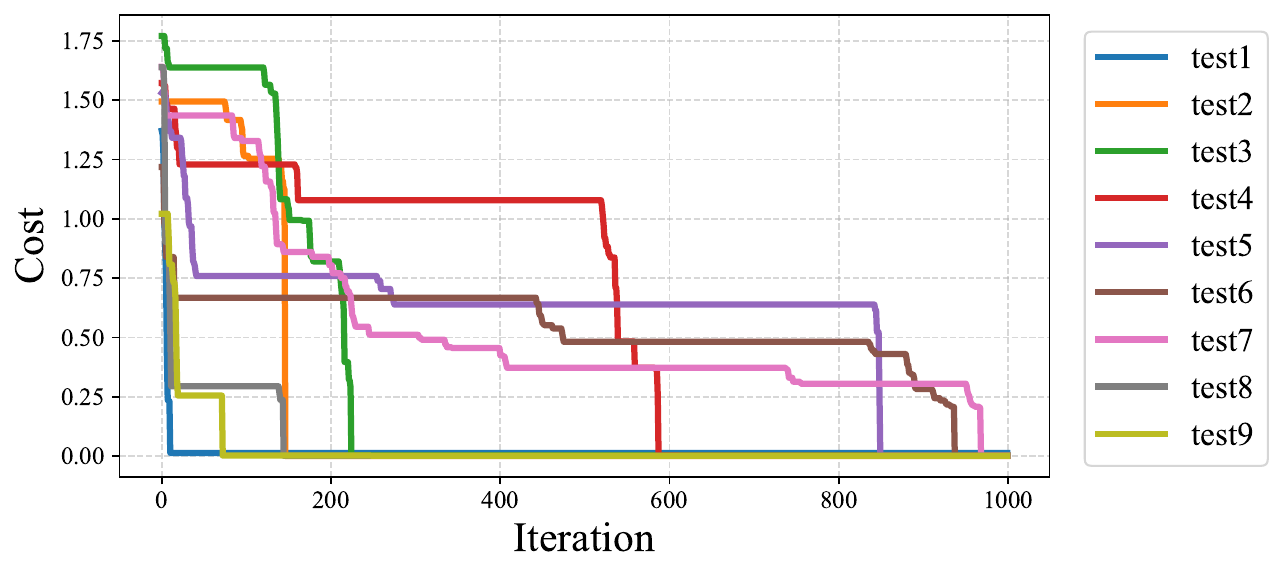}
\vspace{-4ex}
\caption{The cost regression during simulated annealing of ISPD'18 testcases.}
\label{fig:cost}
\end{figure}

In this subsection, multiple experiments are conducted to analyze the insights.
First, the costs of the best boundaries during iterations of Algorithm~\ref{alg:2way} for ISPD'18 testcases are shown in Figure~\ref{fig:cost}. It indicates that the difficulties of finding a high quality solution are different from testcases. However, \myname{} can finally regress to an optimized solution for each testcase.

\begin{figure}[!h]

\centering
\includegraphics[width=0.9\columnwidth]{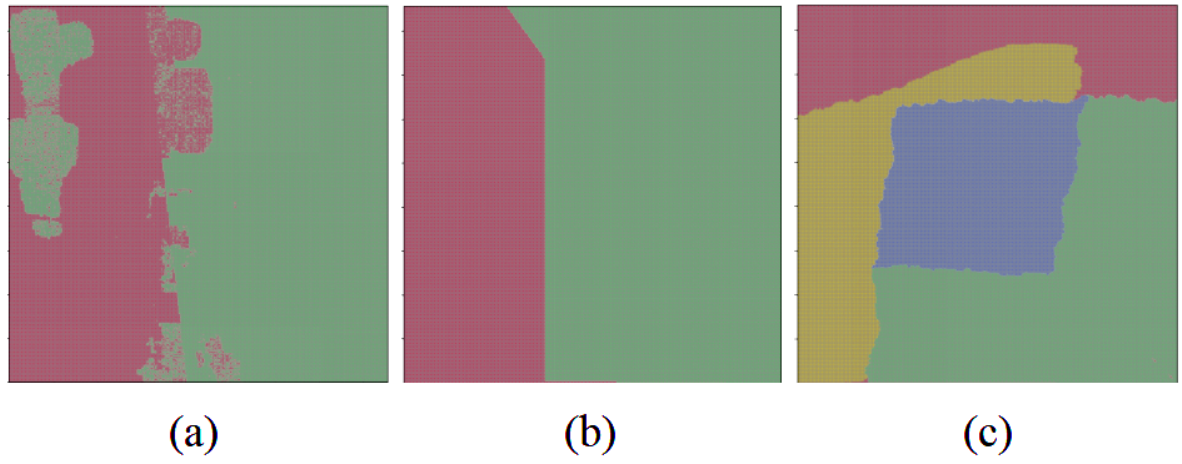}
\caption{The mempool\_tile partitioning of (a) 2-way  from hMETIS; (b) 2-way and (c) 4-way  from \myname{}.}
\label{fig:frag}
\vspace{-2ex}
\end{figure}

Next, some typical example partitioned layouts are in Figure~\ref{fig:frag}. It indicates that hMETIS incurs scattered regions of each partition, resulting in more crossing points on timing critical nets according to Table~\ref{tab:result}, and thus, might degrade the parallel routing quality.
For the 4-way partitioning example, it shows that one partition is located at the center of the layout. This means that although the k-way partitioning is performed by multiple 2-way partitioning, which seems like partitioning in the center is impossible, but with embedding, this kind of solution is also in the search space.

\begin{figure}[!h]
\centering
\includegraphics[width=0.9\columnwidth]{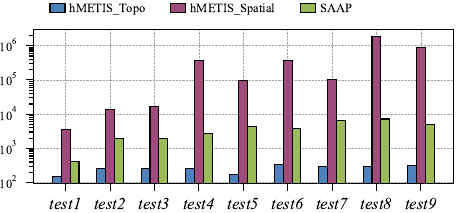}
\caption{The comparisons of topological cut size and spatial cut size of ISPD'18 testcases.}
\label{fig:tospcomp}
\vspace{-2ex}
\end{figure}

Finally, we also show the comparisons of the topological cost of hMETIS with the spatial cut sizes of it and \myname{}, which are in Figure~\ref{fig:tospcomp}. 
This proves that also hMETIS provides the smallest cut sizes topologically, but when mapping the partitioning solutions spatially, the scattering distribution leads to much larger cut sizes. In contrast, \myname{} can provide a reasonable cut size spatially.

\section{Conclusion}
\label{sec:conclude}

In conclusion, this work proposes an analytic spatial-aware partitioning algorithm for parallel routing. With the analytic boundary modeling, regularity-guided simulated annealing, graph embedding, etc., it quickly provides high quality spatially continuous partitions  with small weighted cut sizes and timing-friendly features. 

\bibliographystyle{ACM-Reference-Format}
\bibliography{ref}

\end{document}